\documentclass[a4paper, oneside, 12pt]{article}

\begin{document}

\title{General Classes of Impossible Operations through the
Existence of Incomparable States}

\author{Indrani Chattopadhyay\thanks{ichattopadhyay@yahoo.co.in}
and Debasis Sarkar\thanks{dsappmath@caluniv.ac.in}\\
Department of Applied Mathematics, University of Calcutta,\\
92, A.P.C. Road, Kolkata- 700009, India}

\maketitle

\begin{abstract}

In this work we show that the most general class of anti-unitary
operators are nonphysical in nature through the existence of
incomparable pure bipartite entangled states. It is also shown that
a large class of inner-product-preserving operations defined only on
the three qubits having spin-directions along x, y and z are
impossible. If we perform such an operation locally on a particular
pure bipartite state then it will exactly transform to another pure
bipartite state that is incomparable with the original one. As
subcases of the above results we find the nonphysical nature of
universal exact flipping operation and existence of universal
Hadamard gate. Beyond the information conservation in terms of
entanglement, this work shows how an impossible local operation
evolve with the joint system in a nonphysical way.

\end{abstract}

\section{Introduction}
Quantum systems allow physical operations to perform some tasks that
seems to be impossible in classical domain \cite{mes,mes1,mes2}.
However with the nature of the operations performed it restricts
correctness or exact behavior of the operations to act for the whole
class of states of the quantum system. Possibilities or
impossibilities of various kind of such operations acting on some
specified system is then one of the basic tasks of quantum
information processing. In case of cloning and deleting the input
states must be orthogonal to each other for the exactness of the
operation performed
\cite{wootters,wootters1,wootters2,pati11,pati112}. Rather if the
operation considered is spin-flipping \cite{gisin,unot,unot1} or
Hadamard type then the input set of states enhanced to a great
circle of the Bloch sphere \cite{patili,ghosh,ghosh1}. It indicates
that any angle preserving operation has some restriction on the
allowable input set of states. The unitary nature of all physical
evolution \cite{krause} raised the question that whether the
non-physical nature of the anti-unitary operations is a natural
constraint over the system or not. In other words, it is nice to
show how an impossible operation like anti-unitary, evolve with the
physical systems concerned.

First part of this paper concerns with a connection between general
anti-unitary operations and evolution of a joint system through
local operations together with classical communications, in short
LOCC. Some constraint over the system are always imposed by the
condition that the system is evolved under LOCC. For example,
performing any kind of LOCC on a joint system shared between
distinct parties, the amount of entanglement between some spatially
separated subsystems can not be increased. If we further assume that
the concerned system is pure bipartite, then by Nielsen's criteria
\cite{nielsen,nielsen1} it is possible to determine whether a pure
bipartite state can be transformed to another pure bipartite state
with certainty by LOCC or not. Consequently we find that there are
pairs of pure bipartite states, denoted by incomparable states which
are not interconvertible under LOCC with certainty. The existence of
such class of states prove that the amount of entanglement does not
always determine the possibility of exact transformation of a joint
system by applying LOCC. Now we first pose the problem that would be
discussed in this paper.

Suppose $\rho_{ABCD\ldots}$ be a state shared between distinct
parties situated at distant locations. They are allowed to do local
operations on their subsystems and also they may communicate any
amount of classical information among themselves. But they do not
know whether their local operations are valid physical operations or
not. By valid physical operation we mean a completely positive map
(may be trace-preserving or not) acting on the physical system.
Sometimes an operation is confusing in the sense that it works as a
valid physical operation for a certain class of states but not as a
whole. Therefore they want to judge their local operations using
quantum formalism or other physical principles, may be along with
quantum formalism or may not be. No-signalling, non-increase of
entanglement by LOCC are some of the good detectors of nonphysical
operations \cite{signal,signal1,delsignal,delsignal1,flip,flip1}. In
this paper we want to establish another good detector for a large
number of nonphysical operations. The existence of incomparable
states enables us to find that detector. Suppose $L_A\otimes L_B
\otimes L_C\otimes L_D\otimes \cdots$ be an operation acting on the
physical system represented by $\rho_{ABCD\ldots}$ and
$\rho^{\prime}_{ABCD\ldots}$ be the transformed state. Now it is
known that the states $\rho_{ABCD\ldots}$ and
$\rho^{\prime}_{ABCD\ldots}$ are incomparable by the action of any
deterministic LOCC, then we could certainly say that at least one of
the operations $L_A, L_B, L_C, L_D, \cdots$ are nonphysical.
Therefore if somehow we find two states that are incomparable and by
an operation acting on any party (or a number of parties) one state
is transformed to another then we certainly claim that the operation
is a nonphysical one. We find several classes of nonphysical
operations through this procedure and it is our main motivation in
this work. The paper is organized as follows: in section 2 we
describe what we actually mean by a physical operation and its
relation with LOCC. In section 3 we describe the notion of
incomparability for pure bipartite entangled states. In section 4 we
show the nonphysical nature of the most general class of universal
exact anti-unitary operators through the impossibility of
inter-converting two incomparable states by deterministic LOCC.
Lastly, in section 5 we show a large class of inner-product
preserving operations are also non physical in nature, including the
Hadamard operation. As a subcase of the above operations we
reproduce the nonexistence of exact universal flipping machine
\cite{incomfi}. In all the above cases we have tried to use minimum
number of qubits (only on three spin directions along $x, y, z$) and
the quantum system considered as simple as possible. Also the states
considered here to prove the impossibilities are pure entangled
states.

\section{Physical Operations and LOCC}
In this section we first describe the notion of a physical operation
in the sense of Kraus \cite{krause}. Suppose a physical system is
described by a state $\rho$. By a physical operation on $\rho$ we
mean a completely positive map $\mathcal{E}$ acting on the system
and described by
\begin{equation}
\mathcal{E}(\rho )= \sum_{k} A_{k} \rho A^{\dagger}_{k}
\end{equation}
where each $A_k$ is positive linear operator that satisfies the
relation $ \sum_{k}  A^{\dagger}_{k} A_{k} \leq I$. If $ \sum_{k}
A^{\dagger}_{k} A_{k} = I$, then the operation is trace preserving.
When the state is shared between a number of parties, say, A, B, C,
D,. .... and each $A_k$ has the form $A_k = L^A_k \otimes  L^B_k
\otimes L^C_k \otimes L^D_k \otimes \cdots$ with all the $L^A_k,
L^B_k, L^C_k, L^D_k, \cdots$ are linear positive operators, the
operator is then called a separable superoperator. In this context
we would like to mention an interesting result concerned with LOCC.
Every LOCC is a separable superoperator but it is unknown to us
whether the converse is also true or not. It is further affirmed
that there are separable superoperators which cannot be expressed by
finite LOCC \cite{bennet}. Now if a physical system evolved under
LOCC (may be deterministic or stochastic) then quantum mechanics
does not allow the system to behave arbitrarily. More precisely,
under the action of any LOCC one could find some fundamental
constraints over any entangled system. The content of entanglement
will not increase under LOCC. This is usually known as the principle
of non-increase of entanglement under LOCC. Further for any closed
system as unitarity is the only possible evolution, the constraint
is then: the entanglement content will not change under LOCC. So if
we find some violation of these principles under the action of any
local operation, then we certainly claim that the operation is not a
physical one. No-cloning, no-deleting, no-flipping, all those
theorems are already established with these principles, basically
with the principles of non-increase of entanglement
\cite{flip,flip1}. These kind of proof for those important no-go
theorems will always give us a more powerful physically intuitive
approaches for quantum information processing apart from the
mathematical proofs that the dynamics should be linear as well as
unitary. Linearity and unitarity are the building blocks of every
physical operation \cite{krause,gisin1}. But within the quantum
formalism we always search for better physical situations that are
more useful and intuitive for quantum information processing.
Existence of incomparable states in pure bipartite entangled systems
allow us to use it as a new detector. We have already proved three
impossibilities, viz., exact universal cloning, deleting and
flipping operations by the existence of incomparable states under
LOCC \cite{incomfi,incomfi1} and we would provide some further
classes of nonphysical operations in this paper.

\section{Notion of Incomparability}

To present our work we need to define the condition for a pair of
states to be incomparable with each other. The notion of
incomparability of a pair of bipartite pure states directly follows
from the necessary and sufficient condition for conversion of a pure
bipartite entangled state to another by deterministic LOCC, i.e.,
with probability one. It is prescribed by M. A. Nielsen
\cite{nielsen,nielsen1}. Suppose we want to convert the pure
bipartite state $|\Psi\rangle$ of $d\times d$ system to another
state $|\Phi\rangle$ shared between two parties, say, Alice and Bob
by deterministic LOCC. Consider $|\Psi\rangle$, $|\Phi\rangle$ in
their Schmidt bases $\{|i_A\rangle ,|i_B\rangle \}$ with decreasing
order of Schmidt coefficients: $|\Psi\rangle= \sum_{i=1}^{d}
\sqrt{\alpha_{i}} |i_A i_B\rangle$, $|\Phi\rangle= \sum_{i=1}^{d}
\sqrt{\beta_{i}} |i_A i_B\rangle,$ where $\alpha_{i}\geq
\alpha_{i+1}\geq 0$ and $\beta_{i}\geq \beta_{i+1}\geq0,$ for
$i=1,2,\cdots,d-1,$ and $\sum_{i=1}^{d} \alpha_{i} = 1 =
\sum_{i=1}^{d} \beta_{i}$. The Schmidt vectors corresponding to the
states $|\Psi\rangle$ and $|\Phi\rangle$ are
$\lambda_\Psi\equiv(\alpha_1,\alpha_2,\cdots,\alpha_d),$
$\lambda_\Phi\equiv(\beta_1,\beta_2,\cdots,\beta_d)$. Then Nielsen's
criterion says $|\Psi\rangle\rightarrow| \Phi\rangle$ is possible
with certainty under LOCC if and only if $\lambda_\Psi$ is majorized
by $\lambda_\Phi,$ denoted by $\lambda_\Psi\prec\lambda_\Phi$ and
described as,
\begin{equation}
\begin{array}{lcl}\sum_{i=1}^{k}\alpha_{i}\leq
\sum_{i=1}^{k}\beta_{i}~ ~\forall~ ~k=1,2,\cdots,d
\end{array}
\end{equation}
It is interesting to note that however majorization \cite{maj}
criteria is an algebraic tool, it shows great applicability in
different context of quantum information processing
\cite{majappl,majappl1,majappl2,majappl3}. Now, as a consequence of
non-increase of entanglement by LOCC, if $|\Psi\rangle\rightarrow
|\Phi\rangle$ is possible under LOCC with certainty, then
$E(|\Psi\rangle)\geq E(|\Phi\rangle)$ [where $E(\cdot)$ denote the
von-Neumann entropy of the reduced density operator of any subsystem
and known as the entropy of entanglement]. If the above criterion
(2) does not hold, then it is usually denoted by
$|\Psi\rangle\not\rightarrow |\Phi\rangle$. Though it may happen
that $|\Phi\rangle\rightarrow |\Psi\rangle$ under LOCC. If it
happens that $|\Psi\rangle\not\rightarrow |\Phi\rangle$ and
$|\Phi\rangle\not\rightarrow |\Psi\rangle$ then we denote it as
$|\Psi\rangle\not\leftrightarrow |\Phi\rangle$ and describe
$(|\Psi\rangle, |\Phi\rangle)$ as a pair of incomparable states
\cite{nielsen,incomp}. One of the peculiar feature of such
incomparable pairs is that we are unable to say that which state has
a greater amount of entanglement content than the other. Also for
$2\times 2$ systems there are no pair of pure entangled states which
are incomparable to each other. For our purpose, we now explicitly
mention the criterion of incomparability for a pair of pure
entangled states $|\Psi\rangle, |\Phi\rangle$ of $m\times n$ system
where $\min \{ m,n \}=3$. Suppose the Schmidt vectors corresponding
to the two states are $(a_1, a_2, a_3)$ and $(b_1, b_2, b_3)$
respectively, where $a_1> a_2> a_3~,~b_1> b_2> b_3~,~a_1+ a_2+
a_3=1=b_1+ b_2+ b_3$. In this case the condition for the pair of
states $|\Psi\rangle, |\Phi\rangle$ to be are incomparable to each
other
can be written in the simplified form that\\
\begin{equation}
\begin{array}{lcl}\verb"either," ~ ~ ~ ~a_1 > b_1 ~ ~\verb"and" ~ ~a_3 > b_3\\

\verb"or," ~ ~ ~ ~ ~ ~ ~ ~ ~ ~a_1 < b_1 ~ ~\verb"and" ~ ~a_3 < b_3
\end{array}
\end{equation}
must hold simultaneously.

\section{Incomparability as a Detector for Anti-Unitary Operators}

The general class of anti-unitary operations can be defined in the
form, $\Gamma= CU;$  where $C$ is the conjugation operation and $U$
be the most general type of unitary operation on a qubit, in the
form
$$U=\left(%
\begin{array}{cc}
  \cos\theta & e^{i\alpha}\sin\theta \\
  -e^{i\beta}\sin\theta & e^{i(\alpha+\beta)}\cos\theta \\
\end{array}%
\right)$$

Let us consider three qubit states with the spin-directions along
$x,y,z$ as, $$|0_x
\rangle=\frac{|0\rangle+|1\rangle}{\sqrt{2}}~,~|0_y
\rangle=\frac{|0\rangle+i|1\rangle}{\sqrt{2}}~,
|0_z\rangle=|0\rangle$$

The action of the operator $\Gamma$ on these three states can be
described as,
\begin{quote}
\begin{equation}
\begin{array}{lcl}
\Gamma|0_x\rangle= (\frac{\cos \theta+ e^{-i \alpha}\sin
\theta}{\sqrt{2}})|0\rangle
+e^{-i \beta}(\frac{e^{-i \alpha} \cos\theta-\sin\theta}{\sqrt{2}})|1\rangle , \\
\Gamma|0_y\rangle= (\frac{\cos \theta-i e^{-i \alpha}\sin
\theta}{\sqrt{2}})|0\rangle -e^{-i \beta}(\frac{ie^{-i \alpha}
\cos\theta+\sin\theta}{\sqrt{2}})|1\rangle,
\\ \Gamma |0_z \rangle= \cos \theta |0\rangle -e^{-i \beta} \sin \theta |1\rangle
\end{array}
\end{equation}
\end{quote}

To prove that this operation $\Gamma$ is nonphysical and its
existence leads to an impossibility, we choose a particular pure
bipartite state $|\chi^i \rangle_{AB}$ shared between two spatially
separated parties Alice and Bob in the form,
\begin{equation}
\begin{array}{lcl}|\chi^i\rangle_{AB} & = &\frac{1}{\sqrt{3}} \{|0\rangle_A|0_z
\rangle_B | 0_z \rangle_B + |1 \rangle_A | 0_x \rangle_B | 0_y
\rangle_B\\ & &~ ~+ |2 \rangle_A |0_y \rangle_B |0_x \rangle_B\}
\end{array}
\end{equation}

The impossibility we want to show here is that by the action of
$\Gamma$ locally we are able to convert a pair of incomparable
states deterministically. Now to show incomparability between a pair
of pure bipartite states, the minimum Schmidt rank we require is
three. So the joint state we consider above is a $3\times 4$ state
where Alice has a qutrit and Bob has two qubits. The initial reduced
density matrix of Alice's side is then,
\begin{equation}
\begin{array}{lcl}
\rho_A^i &=&\frac{1}{3}~ \{P[|0\rangle]+P[|1\rangle]+P[|2\rangle]+
\frac{1}{2}(|0\rangle\langle1|+|1\rangle\langle0|\\
& & ~ ~ +|0\rangle\langle2| +|2\rangle\langle0|+|1\rangle\langle2|
+|2\rangle\langle1|) \}
\end{array}
\end{equation}

The Schmidt vector corresponding to the initial state $|\chi^i
\rangle_{AB}$ is $(\frac{2}{3}, \frac{1}{6}, \frac{1}{6} )$.
Assuming that Bob operates $\Gamma$ on one of the two qubits, say
the last one in his subsystem, the joint state shared between Alice
and Bob will transform to
\begin{equation}
\begin{array}{lcl}
|\chi^f \rangle_{AB}&=& \frac{1}{\sqrt{3}} \{|0\rangle_A |0_z
\rangle_B \Gamma(| 0_z \rangle_B) + |1 \rangle_A | 0_x \rangle_B
\Gamma(| 0_y \rangle_B)\\ & & ~ ~+ |2 \rangle_A |0_y \rangle_B
\Gamma(|0_x \rangle_B)\}
\end{array}
\end{equation}

Tracing out Bob's subsystem we again consider the reduced density
matrix of Alice's subsystem. The final reduced density matrix is
\begin{equation}
\begin{array}{lcl}
\rho_A^f &=&\frac{1}{3}~ \{P[|0\rangle]+P[|1\rangle]+P[|2\rangle]+
\frac{1}{2}(|0\rangle\langle1|+|1\rangle\langle0|\\
& & ~ ~+|0\rangle\langle2| +|2\rangle\langle0|-i|1\rangle\langle2|
+i|2\rangle\langle1|) \}
\end{array}
\end{equation}

The Schmidt vector corresponding to the final state $|\chi^f
\rangle_{AB}$ is $(\frac{1}{3}+\frac{1}{2\sqrt{3}}~, \frac{1}{3}~,$
$ \frac{1}{3}-\frac{1}{2\sqrt{3}})$. Interestingly, the Schmidt
vector of the final state does not contain the arbitrary parameters
of the anti-unitary operator $\Gamma$. It is now easy to check that
the final and initial Schmidt vectors are incomparable as,
$\frac{2}{3}~> \frac{1}{3}+\frac{1}{2\sqrt{3}}~> \frac{1}{3}~>
\frac{1}{6}~> \frac{1}{3}-\frac{1}{2\sqrt{3}}$. Thus we have,
$|\chi^i \rangle \not \leftrightarrow |\chi^f \rangle$ so that the
transformation of the pure bipartite state $|\chi^i \rangle$ to
$|\chi^f \rangle$ by LOCC with certainty is not possible following
Nielsen's criteria. Though by applying the anti-unitary operator
$\Gamma$ on Bob's local system the transformation $|\chi^i \rangle
\rightarrow |\chi^f \rangle$ is performed exactly. This
impossibility emerges out of the impossible operation $\Gamma$ which
we have assumed to be exist and apply it to generate the impossible
transformation. Thus we have observed the nonphysical nature of any
anti-unitary operator $\Gamma$ through our detection process. As a
particular case one may verify the non-existence of exact universal
flipper by our method ( choose, $\theta = \pi/2, \alpha = 0, \beta =
0$).

If instead of operating $\Gamma= CU$ we will operate only $U,$
\emph{i.e.}, the general unitary operator, the initial and final
density matrices of one side will be seen to be identical, implying
that there is not even a violation of No-Signalling principle. This
is true as we only operate the unitary operator on any qubit not
restricting on any particular choices, such as they will act
isotropically for all the qubits, etc. Thus it can not
even used to send a signal here.\\

\section{Inner Product Preserving Operations}

In this section we relate the impossibility of some inner product
preserving operations defined only on the minimum number of qubits
$|0_x \rangle, |0_y \rangle, |0_z \rangle$. Here we consider the
existence of the operation defined on these three qubits in the
following manner,
\begin{quote}
\begin{equation}
\begin{array}{lcl}
|0_z \rangle \longrightarrow  (\alpha
|0_z\rangle + \beta |1_z\rangle),\\
|0_x\rangle \longrightarrow  (\alpha
|0_x\rangle + \beta |1_x\rangle), \\
|0_y\rangle \longrightarrow (\alpha |0_y\rangle + \beta
|1_y\rangle),
\end{array}
\end{equation}
\end{quote}
where $|\alpha|^2 + |\beta|^2 =1.$

This operation exactly transforms the input qubit into an arbitrary
superposition on the input qubit with its orthogonal one. To verify
the possibility or impossibility of existence of this operation we
consider a pure bipartite state shared between Alice and Bob:
\begin{equation}
\begin{array}{lcl}|\Pi^i\rangle_{AB}& = & \frac{1}{\sqrt{3}} \{|0\rangle_A (|0_z \rangle
| 0_z \rangle)_B + |1 \rangle_A (| 0_x \rangle | 0_x \rangle)_B
\\ & &~ ~+|2 \rangle_A (|0_y \rangle |0_y \rangle)_B\}
\end{array}
\end{equation}
Reduced density matrix of Alice's side will be of the form,
\begin{equation}
\begin{array}{lcl}
\rho_A^i &=&\frac{1}{3}~ \{P[|0\rangle]+P[|1\rangle]+P[|2\rangle]+
\frac{1}{2}(|0\rangle\langle1|+|1\rangle\langle0|\\
& & ~ ~+|0\rangle\langle2| +|2\rangle\langle0|-i|1\rangle\langle2|
+i|2\rangle\langle1|) \}
\end{array}
\end{equation}

The Schmidt vector corresponding to the initial joint state~
$|\chi^f \rangle_{AB}$ is ~ ~$(\frac{1}{3}+\frac{1}{2\sqrt{3}}~,
\frac{1}{3}~, \frac{1}{3}-\frac{1}{2\sqrt{3}})$. If Bob has a
machine which operates on the three input qubits $|0_x \rangle, |0_y
\rangle, |0_z \rangle$ as defined in equation (9) and he operates
that machine on his local system (say, on the last qubit). Then the
joint state between Alice and Bob will evolve as,
\begin{equation}
\begin{array}{lcl}|\Pi^f \rangle_{AB}&=&
\frac{1}{\sqrt{3}} \{|0\rangle_A |0_z \rangle_B (\alpha |0_z\rangle
+ \beta |1_z\rangle)_B + |1 \rangle_A | 0_x \rangle_B
\\& & (\alpha | 0_x \rangle + \beta |1_x\rangle)_B + |2
\rangle_A |0_y \rangle_B (\alpha|0_y \rangle+ \beta |1_y\rangle)_B\}
\end{array}
\end{equation}
Final reduced density matrix of Alice's side will be of the form,
\begin{equation}
\begin{array}{lcl}
\rho_A^f &=&\frac{1}{3}~ \{P[|0\rangle]+P[|1\rangle]+P[|2\rangle]+
p(|0\rangle\langle1|+|1\rangle\langle0|)\\
& & ~ ~+q|0\rangle\langle2|
+\overline{q}|2\rangle\langle0|+r|1\rangle\langle2|
+\overline{r}|2\rangle\langle1|) \}
\end{array}
\end{equation}where $p=\frac{1}{2}~\{|\alpha|^2-|\beta|^2
+\alpha~\overline{\beta}+\beta~\overline{\alpha}\}$,
~$q=\frac{1}{2}~\{|\alpha|^2+~i|\beta|^2
+\alpha~\overline{\beta}-i\beta~\overline{\alpha}\}$ and
$r=\frac{1}{2}~\{\alpha~\overline{\beta}+\beta~\overline{\alpha}-i\}$.

The eigenvalue equation turns out to be,
$$x^3-(p\overline{p} + q\overline{q} + r\overline{r})x+pr\overline{q}+\overline{pr}q=0,$$
where we denote $1-3\lambda=x.$

To compare the initial and final state we have to check whether the
initial and final eigenvalues will satisfy either of the relations
of equation (3). We rewrite, the above eigenvalue equation as
\begin{equation}
x^3-3Ax+B=0
\end{equation}with, $A=\frac{1}{3}(p\overline{p} + q\overline{q} + r\overline{r})\geq0$ and
$B=pr\overline{q}+\overline{pr}q$. The eigenvalues can then be
written as
$\{\lambda_1\equiv\frac{1}{3}[1-2\sqrt{A}\cos(\frac{2\pi}{3}+\theta)],
~\lambda_2\equiv\frac{1}{3}[1-2\sqrt{A}\cos\theta],~
\lambda_3\equiv\frac{1}{3}[1-2\sqrt{A}\cos(\frac{2\pi}{3}-\theta)]\}$
where $\cos 3\theta=~\frac{-B}{2\sqrt{A^3}}$. We discuss the matter
case by case (for details, see Appendix A).

\textbf{Case-1} : For $\emph{B}<0$, we see an incomparability
between the initial and final joint states if $A= \frac{1}{4}$. In
case $A< \frac{1}{4}$ we observe that either there is an
incomparability between the initial and final states or the
entanglement content of the final state is larger than that of the
initial states. Lastly if $A> \frac{1}{4}$ we also see a case of
incomparability if the condition
$2\sqrt{A}\cos(\frac{2\pi}{3}+\theta)~>~-\frac{\sqrt{3}}{2}$ holds.
Numerical searches support that for real values of $(\alpha,~\beta)$
incomparability is seen almost everywhere in this region.

\textbf{Case-2} : For $\emph{B}=0$, we found that there do not arise
a case of incomparability. It is also seen that there is always an
increase of entanglement by LOCC if $A< \frac{1}{4}$, which is the
only possibility for real values of $\alpha,\beta$.

\textbf{Case-3} : For $\emph{B}>0$ we also get a similar result like
Case-1. Only the condition for incomparability in case $A>
\frac{1}{4}$ if changed to the form that $2\sqrt{A}\cos\varphi <
\frac{\sqrt{3}}{2}$ where
$\varphi=\min\{\theta,~(\frac{2\pi}{3}-\theta)\}\in(\frac{\pi}{6},~\frac{\pi}{3}).$
It must be noted that for real values of $\alpha,\beta$ this subcase
do not arise at all.

In particular if we check the values of $\alpha, \beta$ be such that
they represents the operations flipping(i.e., $\alpha = 0$) and
Hadamard(i.e., $\alpha = \beta = \frac{1}{\sqrt{2}}$) respectively,
we find from the above that in both the cases the initial and final
states are incomparable.

Thus we get almost in all cases some kind of violation of physical
laws implying that the kind of inner product preserving operations
defined  on only three states is nonphysical in nature and we
observe for a large class of such inner-product-preserving operation
incomparability senses.

To conclude this work proves a close relation between anti-unitary
operators and the existence of incomparable states. Incomparability
shows it is also able to detect nonphysical operations like Hadamard
and some other inner-product preserving operations. This work also
shows an interplay between LOCC, nonphysical operations and the
entanglement behavior of quantum systems.

{\bf Acknowledgements:}  We would like to thank the referee for
valuable suggestions and useful comments. The authors are grateful
to Dr. A. K. Pati and Dr. P. Agarwal for useful discussions
regarding this work. I.C. also acknowledges CSIR, India for
providing fellowship during this work.

\appendix

\textbf{Case-1 : $\emph{B}<0$} This implies $3\theta \in
[0,~\frac{\pi}{2}) \bigcup (\frac{3\pi}{2},~2\pi].$ We analyze this
in two section.

If $3\theta \in [0,\frac{\pi}{2})$ we have, $\frac{\sqrt{3}}{2}~<~
\cos\theta~
\leq~1~\Rightarrow~\lambda_2~\in~[\frac{1}{3}(1-2\sqrt{A}),\frac{1}{3}(1-\sqrt{3A}))$.
Again $0~\leq~\theta~<~\frac{\pi}{6}~
\Rightarrow~-\frac{\sqrt{3}}{2} < \cos(\frac{2\pi}{3}+\theta) \leq
-\frac{1}{2} ~\Rightarrow
\lambda_1~\in~[\frac{1}{3}(1+\sqrt{A}),\frac{1}{3}(1+\sqrt{3A}))$.
Finally, $0~\leq~\theta~<~\frac{\pi}{6}~\Rightarrow~
\cos(\frac{2\pi}{3}-\theta)\in [-\frac{1}{2},0) ~\Rightarrow
\lambda_3~\in~(\frac{1}{3},\frac{1}{3}(1+\sqrt{A})]$.

Otherwise $~3\theta \in (\frac{3\pi}{2},2\pi],$ i.e., $\theta \in
(\frac{\pi}{2},\frac{2\pi}{3}]$, we have,
$\lambda_3~\in~[\frac{1}{3}(1-2\sqrt{A}),\frac{1}{3}(1-\sqrt{3A}))$,
$\lambda_2~\in~[\frac{1}{3}(1+\sqrt{A}),\frac{1}{3}(1+\sqrt{3A}))$
and $\lambda_1~\in~(\frac{1}{3},\frac{1}{3}(1+\sqrt{A})]$.

Thus in both the cases
$\lambda^f_{MAX}~\in~[\frac{1}{3}(1+\sqrt{A}),
\frac{1}{3}(1+\sqrt{3A}))$
and \\$\lambda^f_{MIN} ~\in~ [\frac{1}{3}(1-2\sqrt{A}), \frac{1}{3}(1-\sqrt{3A}))$.\\

For $A=\frac{1}{4}$ we observe that
${\lambda^f}_{MIN}\in[0,\frac{1}{3}(1-\frac{\sqrt{3}}{2}))<{\lambda^i}_{MIN}$
and
${\lambda^f}_{MAX}\in[\frac{1}{2},\frac{1}{3}(1+\frac{\sqrt{3}}{2}))<{\lambda^i}_{MAX}$
which implies that $|\Pi^i\rangle_{AB}~,|\Pi^f\rangle_{AB}$ are
incomparable.

If $A~<~\frac{1}{4}$ then ${\lambda^f}_{MAX}\leq {\lambda^i}_{MAX}$.
So, in case ${\lambda^f}_{MIN}\leq {\lambda^i}_{MIN}$ the states
$|\Pi^i\rangle_{AB}~,~|\Pi^f\rangle_{AB}$ are incomparable,
otherwise we have ${\lambda^f}_{MIN}\geq {\lambda^i}_{MIN}$  then
$E(|\Pi^i\rangle_{AB})~<~E(|\Pi^f\rangle_{AB})$. For real values of
$\alpha,~ \beta$ we can express $A$,$B$ as
\begin{equation}
\begin{array}{lcl}
~ ~\emph{A}= \frac{1}{4}+ \frac{1}{6}[ 2{\alpha}^2{\beta}^2+
3\alpha\beta({\alpha}^2-{\beta}^2)] \\
\emph{B}= \frac{\beta}{4} ({\alpha}^2-{\beta}^2+ 2\alpha\beta)
[\alpha(2{\alpha}^2+1)+ \beta({\alpha}^2-{\beta}^2)]
\end{array}
\end{equation}
Numerical evidences support that for real $\alpha$, $\beta$ most of
the cases show incomparability between
$|\Pi^i\rangle_{AB}~,~|\Pi^f\rangle_{AB}$.

Lastly if $A~\geq~\frac{1}{4}$ then
${\lambda^i}_{MAX}~<~{\lambda^f}_{MAX}$. Thus incomparability
between $|\Pi^i\rangle_{AB}~,~|\Pi^f\rangle_{AB}$ will hold if
${\lambda^i}_{MIN}~<~{\lambda^f}_{MIN}$. For this we get the
condition that $2\sqrt{A}\cos\phi~<~\frac{\sqrt{3}}{2}$ where
$\phi=\min\{\theta,~\frac{2\pi}{3}-\theta\}\in(\frac{\pi}{6},\frac{\pi}{3})$.
For real values of $\alpha,\beta$ from equation(A.1), we see
$A>\frac{1}{4}$ implies $B>0$. Thus for real $\alpha,\beta$ this subcase do not arises.\\

{\textbf{Case-2 : $B=0.$} Here the final eigenvalues are
$\{\frac{1}{3}(1+\sqrt{3A}),
~\frac{1}{3},~\frac{1}{3}(1-\sqrt{3A})\}.$ Thus,
$E(|\Pi^i\rangle_{AB})~\geq~E(|\Pi^f\rangle_{AB})$  if $A~\geq
\frac{1}{4}$. Incomparability between the initial and final joint
states $|\Pi^i\rangle_{AB}~,|\Pi^f\rangle_{AB}$ will not occur in
this case.

Hence for all values of $\alpha,\beta$ for which $A<\frac{1}{4}$
there is an increase of entanglement by applying the local operation
defined in equation(9) in Bob's system. This impossibility indicate
the impossibility of the operation defined in (9) for those values
of $\alpha,\beta$ which satisfy $A~<~\frac{1}{4}$. And for real
values of $\alpha, \beta$, in all possibilities for $\emph{B}=0$
we have $\emph{A}~<~\frac{1}{4}$.  This case always shows an increase of entanglement.\\

\textbf{Case-3 : $\emph{B}>0$.} Here $3\theta \in
(\frac{\pi}{2}~,\frac{3\pi}{2}) \Rightarrow \theta \in
(\frac{\pi}{6}~,\frac{\pi}{2}) ~\Rightarrow \cos \theta \in
(\frac{\sqrt{3}}{2}~ ,0)\Rightarrow
\lambda_2\in(\frac{1}{3}(1-\sqrt{3A}),\frac{1}{3})$. Again $\theta
\in (\frac{\pi}{6}~,\frac{\pi}{2})\Rightarrow
\cos(\frac{2\pi}{3}+\theta)\in (-1,-\frac{\sqrt{3}}{2})
\Rightarrow\lambda_1\in(\frac{1}{3}(1+\sqrt{3A})~,\frac{1}{3}(1+2\sqrt{A}))$.
Lastly, $\theta \in (\frac{\pi}{6},\frac{\pi}{2}) ~\Rightarrow
\cos(\frac{2\pi}{3}-\theta)\in(0,\frac{\sqrt{3}}{2})~\Rightarrow
\lambda_3\in (\frac{1}{3}(1-\sqrt{3A}),\frac{1}{3})$.

Hence in this case
$\lambda^f_{MAX}~\in~(\frac{1}{3}(1+\sqrt{3A}),\frac{1}{3}(1+2\sqrt{A}))$
and $\lambda^f_{MIN} ~\in
(\frac{1}{3}(1-\sqrt{3A}),\frac{1}{3})$.\\
So for $A~=~\frac{1}{4}$ we have
${\lambda^i}_{MAX}~<~{\lambda^f}_{MAX}$ and
${\lambda^i}_{MIN}~<~{\lambda^f}_{MIN}$ implies that
$|\Pi^i\rangle_{AB}~,~|\Pi^f\rangle_{AB}$ are incomparable.

Again for $A~\leq~\frac{1}{4}$ we see,
${\lambda^f}_{MIN}~>~{\lambda^i}_{MIN}$. Thus if
${\lambda^f}_{MAX}~>~{\lambda^i}_{MAX}$ then the states
$|\Pi^i\rangle_{AB}~,~|\Pi^f\rangle_{AB}$ are incomparable or if
${\lambda^f}_{MAX}~<~{\lambda^i}_{MAX}$ then
$E(|\Pi^i\rangle_{AB})~<~E(|\Pi^f\rangle_{AB})$.

Lastly if $A~\geq~\frac{1}{4}$ then
${\lambda^i}_{MAX}~<~{\lambda^f}_{MAX}$. Incomparability between the
initial and final joint states
$|\Pi^i\rangle_{AB}~,~|\Pi^f\rangle_{AB}$ will hold if
${\lambda^i}_{MIN}~<~{\lambda^f}_{MIN}$. For this we get the
condition that
$2\sqrt{A}\cos(\frac{2\pi}{3}+\theta)~>~-\frac{\sqrt{3}}{2}$. From
equation (A.1) we find, for real values of $\alpha$ and $\beta$,
numerical results support that in most of cases there is an
incomparability between $|\Pi^i\rangle_{AB}~,~|\Pi^f\rangle_{AB}$.

\end{document}